\documentclass[12pt]{article}

\usepackage{hyperref}

\usepackage{appendix}
\usepackage{amsmath, amssymb, mathrsfs}
\usepackage{tikz, tcolorbox, color}
\usetikzlibrary{decorations.markings, arrows.meta, calc}
\tcbuselibrary{skins}
\usepackage {diagbox}
\usepackage{ulem}
\usepackage{physics}
\usepackage{fancybox}
\usepackage{fancyhdr}
\usepackage{bm}

\usepackage{epsf}
\usepackage{amsmath,amssymb}
\usepackage{latexsym}
\usepackage{graphicx,color}
\usepackage{wrapfig}
\usepackage{latexsym}
\usepackage{graphics}

\usepackage{color}
\usepackage{delarray,color} 

\usepackage[T1]{fontenc}
\usepackage{lmodern}
\usepackage[utf8]{inputenc}

\usepackage{mathtools}

\newcommand {\beq}{\begin{align}}
\newcommand {\eeq}{\end{align}}

\newcommand{\be}{\begin{equation}}
\newcommand{\ba}{\begin{align}}
\newcommand{\ea}{\end{align}}
\newcommand{\ee}{\end{equation}}

\newcommand{\beqa}{\begin{align}}
\newcommand{\eeqa}{\end{align}}

\newcommand{\unit}{\hbox to 3.8pt{\hskip1.3pt \vrule height 7.4pt
    width .4pt \hskip.7pt \vrule height 7.85pt width .4pt \kern-2.4pt
    \hrulefill \kern-3pt \raise 3.7pt\hbox{\char'40}}}

\def\matt[#1,#2,#3,#4]{\left(%
\begin{array}{cc} #1 & #2 \\ #3 & #4 \end{array} \right)}

\usetikzlibrary{arrows.meta,positioning,calc,decorations.pathmorphing,patterns}

\usepackage{tabularx}

\catcode`\@=11
\@addtoreset{equation}{section}

\catcode`@=12
\relax

\DeclareMathOperator{\End}{End}
\DeclareMathOperator{\Alg}{Alg}

\newcommand{\Hfull}{\mathcal H}
\newcommand{\Hlow}{\mathcal H_{\rm low}}
\newcommand{\Hphys}{\mathcal H_{\rm phys}}

\newcommand{\Hlogical}{\mathcal H_{\rm logical}}
\newcommand{\A}{\mathcal A}
\newcommand{\Rcp}{\mathfrak R}
\newcommand{\Lcp}{\mathfrak L_{\rm cp}}
\newcommand{\Ccom}{\mathfrak C_{\rm com}}
\newcommand{\one}{\mathbf 1}

\newtheorem{definition}{Definition}[section]

\newtheorem{lemma}[definition]{Lemma}

\begin{document}

\begin{titlepage}

\setcounter{page}{0}

\renewcommand{\thefootnote}{\fnsymbol{footnote}}

\begin{flushright}

YITP-26-88

\end{flushright}

\vskip 1.35cm

\begin{center}
{\Large \bf 
Entanglement Wedge Reconstruction without\\
Holographic Quantum Error Correction
}

\vskip 1.2cm 

{\normalsize
Seiji Terashima${}^{a}$~\footnote{terasima(at)yukawa.kyoto-u.ac.jp}
}

\vskip 0.8cm


${}^{a}${\it
Center for Gravitational Physics and Quantum Information, \\
Yukawa Institute for Theoretical Physics, Kyoto University, Kyoto 606-8502, Japan
}

\end{center}

\vspace{12mm}

\begin{abstract}

Bulk reconstruction is a central problem in AdS/CFT, and entanglement
wedge reconstruction is its subregion version.  We argue that this
subregion statement should be separated from the stronger holographic
quantum error correction interpretation, in which one
region-independent logical bulk operator has code-preserving
representatives in several boundary regions.  A simple locality argument
shows that such a common reconstruction must commute with the
code-preserving local algebras in the complementary regions.  This is
the mechanism realized in HaPPY-type codes: the erased regions are blind
to a protected logical algebra.  An ordinary finite \(N\) holographic
CFT does not have such a protected invisible sector for supergravity fields.
Its low-energy local observables, in particular, suitably smeared
stress tensors, detect the physical support and gravitational dressing
of ordinary bulk operators, up to possible center or superselection
data.  
Thus, there is no such holographic
quantum error correction and the \(N=\infty\) agreement of global and subregion HKLL
formulae is a free-theory statement.  What remains is entanglement wedge reconstruction without holographic
quantum error correction, or subregion complementarity: each boundary region has its own code-preserving low-energy algebra and
its own region-adapted bulk interpretation, rather than a shared logical
operator.

\end{abstract}

\end{titlepage}

\newpage

\tableofcontents
\vskip 1.2cm

\setcounter{footnote}{0}

\section{Introduction and summary}

Bulk reconstruction is one of the central questions in AdS/CFT correspondence \cite{Maldacena:1997re}.  
The duality should not only reproduce boundary correlation functions \cite{Gubser:1998bc, Witten:1998qj} but also explain how semiclassical bulk observables are represented in the conformal field theory (CFT).  
In the bulk construction \cite{Banks:1998dd, Bena:1999jv, Duetsch:2002hc, Hamilton:2006az, Hamilton:2005ju, Balasubramanian:1998sn, Heemskerk:2012np}, a bulk field in a fixed semiclassical background is reconstructed from CFT operators on the full boundary, at least in the large $N$ effective field theory regime.

Entanglement wedge reconstruction (EWR) \cite{Dong:2016eik, Harlow:2016vwg} is a proposed subregion version of this idea \footnote{
An explicit example of this is the AdS-Rindler bulk reconstruction \cite{Hamilton:2006az}.
}, motivated by the JLMS relation \cite{Jafferis:2015del}.  
Given a boundary region $A$, one asks whether bulk observables in the corresponding entanglement wedge, bounded by the asymptotic boundary and the (generalized) Ryu-Takayanagi surface \cite{Ryu:2006bv, Lewkowycz:2013nqa}, can be represented by CFT operators supported only in $A$.  

Entanglement wedge reconstruction is often explained as a manifestation
of holographic quantum error correction (HQEC).  In that explanation, a bulk
operator acting on a low-energy code subspace has several boundary
representatives, supported on different authorized boundary regions, and
all of them act as the same logical operator.  This idea was introduced
in AdS/CFT context in \cite{Almheiri:2014lwa} to resolve the ``paradox'' of bulk locality, realized in toy tensor
network models such as the HaPPY code \cite{Pastawski:2015qua}, and
formulated algebraically in relation to entanglement wedge reconstruction
and the Ryu--Takayanagi formula \cite{Dong:2016eik,Harlow:2016vwg}.

The purpose of this paper is to separate this strong error-correcting
claim from a weaker, region-by-region form of reconstruction.  We call
the strong claim EWR with HQEC.  It says that there is a
region-independent logical algebra, and that the same element of this
logical algebra has code-preserving representatives in several different
boundary regions.  By contrast, EWR without HQEC says the following.  For
each boundary region $A$, one considers the CFT operators
supported in $A$ that preserve the chosen low-energy sector $\Hlow$, and
then restricts them to $\Hlow$.  The resulting algebra, denoted
$\Rcp(A)$ below, may be interpreted semiclassically as a bulk algebra
with gravitational dressing compatible with $A$.  No equality between the
operators associated with different boundary regions is assumed.

We give a simple algebraic explanation of why the stronger HQEC
interpretation fails.
If a low-energy operator $w$ has 
code-preserving representatives in several regions $R_i$, locality
implies that $w$ commutes, on $\Hlow$, with every code-preserving local
algebra supported in the complementary region $R_i^c$.
In an ordinary local
CFT/QFT, low-energy local observables such as smeared stress tensors and
other single-trace operators are expected to detect ordinary bulk
excitations.  
Their commutant contains no ordinary bulk-field operator,
up to possible center or superselection data.
With this requirement from physics, such a common reconstruction $w$ should be trivial.
HaPPY-type codes evade
this conclusion because this basic requirement is not satisfied.
This conclusion was given in \cite{Terashima:2020uqu, Terashima:2021klf, Sugishita:2022ldv, Sugishita:2023wjm,  Sugishita:2024lee} based on 
explicit computations in \cite{Terashima:2017gmc, Terashima:2023mcr}
and the discussions in this paper may be regarded as a simple and algebraic explanation of them.

This argument is not a no-go theorem for subregion reconstruction itself.
It rules out only the stronger common-logical-operator interpretation
associated with HQEC.  What remains is EWR without HQEC: for each
boundary region $A$, the code-preserving low-energy algebra
$\Rcp(A)$ may be given an $A$-dependent bulk interpretation
as an algebra of gauge-invariant operators whose gravitational dressing
is compatible with $A$.  The important point is that the regional
algebras for different boundary regions need not be different
representations of one region-independent logical algebra.

This distinction is essential in gravity.  At finite $N$, a local bulk
field is not by itself a gauge-invariant operator; a physical bulk
operator must include gravitational dressing
\cite{Donnelly:2015hta, Donnelly:2016rvo}.  Different choices of boundary
support generally lead to different dressings, and hence to different
physical CFT operators.  Thus, two region-adapted reconstruction formulae
may give equivalent leading semiclassical descriptions in an overlapping
classical region, while failing to define the same finite $N$ operator.
This is the operator-theoretic content of subregion complementarity
\cite{Sugishita:2023wjm, Sugishita:2024lee}.

Note that the strict $N=\infty$ limit does not remove this issue.  In that limit
the relevant holographic sector is a generalized free field (GFF) theory,
equivalent to a free bulk theory.  Global HKLL reconstruction and
AdS--Rindler HKLL reconstruction can then be identified as two
descriptions of the same free field \cite{Hamilton:2006az}.  However,
this is a free-theory identity, not evidence for a nontrivial finite $N$
error correcting code.  The GFF sector does not have the
full local operator structure of the interacting CFT, in particular, the
finite $N$ stress tensor.  Previous analyses showed that global and
subregion reconstructions that agree in the strict free limit become
distinguishable at the first nontrivial interaction or gravitational
order, even within low-energy matrix elements
\cite{Sugishita:2022ldv,Sugishita:2023wjm,Sugishita:2024lee}.

The paper is organized as follows.  Section~\ref{sec:two-notions}
defines EWR with HQEC and EWR without HQEC.  Section~\ref{sec:finite-dimensional}
proves the finite-dimensional commutant lemma and explains its
three-region and two-region consequences.  Section~\ref{sec:gff}
explains why the strict $N=\infty$ generalized-free limit is not evidence
for HQEC at finite $N$, and discusses the role of approximate recovery.
Section~\ref{s5} gives the AdS/CFT interpretation as EWR without HQEC. 

\section{EWR with and without HQEC}
\label{sec:two-notions}

In discussions of subregion reconstruction in AdS/CFT, two logically
different statements are often conflated.  The first is an HQEC
interpretation of EWR \cite{Almheiri:2014lwa, Dong:2016eik, Harlow:2016vwg, Cotler:2017erl}: a single logical bulk operator is assumed to have
boundary representatives in several different regions.  The second is a region-by-region reconstruction statement: for each boundary
region \(A\), the regional low-energy CFT algebra \(\Rcp(A)\) may be
given an \(A\)-dependent bulk interpretation as an algebra of
gauge-invariant, gravitationally dressed operators.  This second
statement does not require the same logical operator to be represented
in different boundary regions \cite{Sugishita:2023wjm, Sugishita:2024lee, Terashima:2020uqu, Terashima:2021klf}.

Throughout this paper $\A(A)$ denotes the CFT operator algebra supported in the boundary region $A$ on the full Hilbert space, and $P$ denotes the projector onto the chosen low-energy sector $\Hlow$.\footnote{
This low energy sector $\Hlow$ is what is commonly called the code subspace. The definition of this is known to have some subtleties, which we will not discuss.
}  
We define the following subalgebra of $\End(\Hlow)$
\begin{equation}
 \Rcp(A):=
 \{O|_{\Hlow}:O\in\A(A),\ [O,P]=0\}.
 \label{eq:Rcp-section2}
\end{equation}
This is a boundary-side object, which will be regarded as the bulk subalgebra corresponding to the subregion $A$.  
For the algebraic version of the EWR, this algebra is interpreted as the algebra of gauge-invariant bulk operators whose gravitational dressing is compatible with the region $A$.
It is not the full UV algebra $\A(A)$.  It is also not obtained by simply taking the projection of an arbitrary operator $O$ in $\A(A)$, $POP$: although $POP$ always preserves $\Hlow$ and $[POP, P]=0$, it need not be an operator in $\A(A)$ because the projector $P$ is global.

The bulk operator $o\in \End(\Hlow)$ is said to be reconstructible from two boundary regions $A$ and $B$ if there exist operators
\begin{equation}
 O_A\in\A(A),\qquad O_B\in\A(B)
\end{equation}
which preserve the low-energy sector and give the same action on it as $o$:
\begin{equation}
 [O_A,P]=[O_B,P]=0,
 \qquad
 O_A|_{\Hlow}=O_B|_{\Hlow}=o.
 \label{eq:hqec-two}
\end{equation}

\subsection{EWR with HQEC}

EWR with HQEC, in the ADH/HaPPY interpretation \cite{Almheiri:2014lwa, Pastawski:2015qua} emphasized here, is the
statement that there is a region-independent logical algebra whose same
elements have representatives in different boundary regions.  This
interpretation is closely related to the operator-algebra formulations of
entanglement wedge reconstruction in \cite{Dong:2016eik,Harlow:2016vwg},
although the present paper focuses on the common-logical-operator aspect.
In its simplest factorized version \cite{Dong:2016eik}, one writes, approximately,
\begin{equation}
 \Hlow\simeq \mathcal H_a\otimes\mathcal H_{\bar a},
\end{equation}
and the subalgebra $\Rcp(A)$ is $\End(\mathcal H_a)\otimes \one$.  Operator-algebra versions \cite{Harlow:2016vwg} 
do not assume this factorization of the bulk Hilbert space.
In two-region language, HQEC claims possible extra common reconstructions:
\begin{equation}
 \Rcp(A)\cap\Rcp(B)
 \supsetneq
 \Rcp(A\cap B),
 \label{eq:extra-overlap-section2}
\end{equation}
which means that some bulk operators are reconstructible from both $A$ and $B$ even though they are not reconstructible from $A \cap B$.\footnote{
This follows from the observation that the global and AdS-Rindler HKLL bulk reconstructions yield the same bulk local operator in an overlapping classical region
\cite{Almheiri:2014lwa}; however, this holds only in the $N=\infty$ limit \cite{Terashima:2020uqu, Sugishita:2023wjm} because of the gravitational dressing in the bulk picture
in the algebraic version \cite{Sugishita:2024lee}.
}
In a three-region secret-sharing setup, the corresponding signature is a nontrivial operator that is reconstructible from each authorized pair,
while its action is not detected by the one-region code-preserving local algebra of any single
region.

\subsection{EWR without HQEC}

EWR without HQEC is the algebraic version of the EWR, but it does not assume \eqref{eq:extra-overlap-section2} and essentially gives
\begin{equation}
 \Rcp(A)\cap\Rcp(B)
 =
 \Rcp(A\cap B).
 \label{aandb}
\end{equation}

A different region $B$ has its own algebra $\Rcp(B)$ and its own bulk interpretation. A local-looking field, which appears in the $N=\infty$ limit, in the overlap of two classical wedges may have an $A$-dressed version and a $B$-dressed version.  These can agree at the strict free limit but differ when we include $1/N$ corrections and in the finite $N$ setting.

The term ``weak EWR'' was used in earlier works \cite{Terashima:2020uqu, Sugishita:2023wjm, Sugishita:2024lee} to denote the second EWR without HQEC, in contrast to the EWR with the factorized Hilbert space \cite{Dong:2016eik}.  
In the present paper, we refer to it as EWR without HQEC.  The point is not that the HQEC equality holds only approximately, but that the HQEC equality is not the right statement to impose.  What remains is a regional interpretation of low-energy CFT algebras as dressed gravitational algebras.

The rest of the paper analyzes why the HQEC equality fails, and what remains once it is replaced by EWR without HQEC.

\section{Finite-dimensional formulation}
\label{sec:finite-dimensional}

We now consider regulated finite-dimensional models and
show that there is no common reconstruction, i.e., no HQEC structure, by a simple lemma combined with a requirement from quantum field theory that is absent in purely quantum-information models.
This abstracts the
locality part of the three-region energy-momentum-tensor argument of
\cite{Sugishita:2024lee, Sugishita:2023wjm}.  
The lemma itself is a statement of quantum information and local operator algebras: a common reconstruction must commute
with the code-preserving local algebras in the complementary regions.  To apply it to AdS/CFT, one uses the ordinary quantum field theory fact that 
sufficiently many
low-energy local observables, such as smeared stress tensors, do not leave a
nontrivial ordinary low-energy operator in their commutant.

\subsection{Local algebras and bulk reconstructions}

We now introduce a finite-dimensional toy model that abstracts the
algebraic issue of the bulk reconstruction in AdS/CFT.  
In this section \(\Hfull\) denotes a finite-dimensional
toy-model Hilbert space.\footnote{
The toy model is not meant to be a finite-dimensional CFT, of course.
}
To each allowed spatial region $X$ associate a unital $*$-subalgebra
\begin{equation}
 \A(X)\subset\End(\Hfull).
\end{equation}
Here, for the finite-dimensional model, the Hilbert space is the tensor product of the Hilbert spaces associated with the points of a discretized space.
We assume the inclusion property and locality:
\begin{align}
 X\subset Y&\quad\Longrightarrow\quad \A(X)\subset\A(Y),\label{eq:isotony}\\
 X\cap Y=\varnothing&\quad\Longrightarrow\quad [\A(X),\A(Y)]=0,\label{eq:locality}
\end{align}
which should be satisfied for quantum field theories.

Let \(P\) be the projector onto a distinguished
subspace
\begin{align}
  \Hlow\subset \Hfull,   
\end{align}
which corresponds to the low-energy supergravity subspace around the background, or the code subspace in the usual terminology.
Now we define the subalgebra which corresponds to the bulk algebra reconstructed from the CFT operators supported on $X$. 
\begin{definition}[bulk reconstruction algebra]
For a region $X$, define
\begin{equation}
 \Rcp(X):=
 \left\{
 x|_{\Hlow}:x\in\A(X),\ [x,P]=0
 \right\}
 \subset\End(\Hlow).
 \label{eq:Rcp}
\end{equation}
\end{definition}

The algebra $\Rcp(X)$ contains physical representatives that act intrinsically within the chosen low-energy theory. \footnote{
This definition is the same as the usual one used in the holographic error correction code, as explained 
in Appendix \ref{app:code-preservation}
} 
In general, it is smaller than the projection of $\A(X)$, 
\begin{equation}
 P\A(X)P|_{\Hlow}.
 \label{eq:compression-set}
\end{equation}

\subsection{Derivation of no HQEC}

Let $R_1,\ldots,R_m$ be subregions.
Define the common reconstruction algebra
\begin{equation}
 \Ccom(R_1,\ldots,R_m)
 :=\bigcap_{i=1}^m\Rcp(R_i),
 \label{eq:common-algebra}
\end{equation}
which consists of the bulk operators reconstructible from every $R_i$, 
and the complementary algebra 
\begin{equation}
 \Lcp(R_1^c,\ldots,R_m^c)
 :=\bigvee_{i=1}^m\Rcp(R_i^c)
 =\Alg_{\Hlow}\left(\bigcup_{i=1}^m\Rcp(R_i^c)\right).
 \label{eq:complementary-cp-algebra}
\end{equation}

\begin{lemma}[Common reconstructions are locally invisible]
\label{lem:commutant-bound}
Under physical locality \eqref{eq:locality},
\begin{equation}
 \Ccom(R_1,\ldots,R_m)
 \subseteq
 \Lcp(R_1^c,\ldots,R_m^c)'
 \label{eq:commutant-bound}
\end{equation}
where the commutant is taken in $\End(\Hlow)$.
\end{lemma}

Let $w\in\Ccom(R_1,\ldots,R_m)$.  For each $i$ there is an operator
\begin{equation}
 \widetilde w_i\in\A(R_i),
 \qquad [\widetilde w_i,P]=0,
 \qquad \widetilde w_i|_{\Hlow}=w.
 \label{eq:wi-rep}
\end{equation}
Take any $x\in\Rcp(R_i^c)$.  By definition, there is an operator
\begin{equation}
 \widetilde x_i\in\A(R_i^c),
 \qquad [\widetilde x_i,P]=0,
 \qquad \widetilde x_i|_{\Hlow}=x.
\end{equation}
Since $R_i$ and $R_i^c$ are disjoint, physical locality gives
\begin{equation}
 [\widetilde w_i,\widetilde x_i]=0.
\end{equation}
Because both representatives preserve $\Hlow$, this descends to
\begin{equation}
 [w,x]=0
 \qquad\text{on }\Hlow .
 \label{eq:code-preserving-commutator}
\end{equation}
Thus $w$ commutes with $\Rcp(R_i^c)$ for every $i$, and hence with the algebra generated by all complementary algebras.

Equation~\eqref{eq:commutant-bound} is the unconditional information-theoretic statement.

\subsubsection{Local QFT detectability }
\label{subsec:local-qft-detectability}

The additional physical input is simple: in an ordinary QFT, local low-energy
observables, in particular, suitably smeared stress tensors, distinguish
ordinary low-energy excitations.
Low-energy (almost) local observables, including suitably smeared stress tensors and other low-energy single-trace observables, do not leave an ordinary low-energy operator in their commutant (at least, on a low-energy sector of an ordinary CFT).
Here, we assume that such observables are available for the whole space.
If an operator commutes with all such observables, it is at most a central, topological, or superselection datum. 
None of these possibilities is the focus of the bulk reconstruction here, because what HQEC requires is the usual bulk operators.
By this, we ignore these possible non-bulk-field operators for notational simplicity; then, from this QFT property, we have
\begin{equation}
\Lcp(R_1^c,\ldots,R_m^c)'=\mathbb C\one_{\Hlow},
 \label{eq:LCI}
\end{equation}
if $\bigcup_{i=1}^m R_i^c$ is the whole space, which is equivalent to $\bigcap_{i=1}^m R_i=\varnothing$,
because any low-energy (almost) local observable may be included in $\Rcp(R_i^c)$, where $i$ is chosen from  $\{ 1, \dots, m \}$.
Together with Lemma~\ref{lem:commutant-bound}, this condition implies
\begin{equation}
 \Ccom(R_1,\ldots,R_m)=\mathbb C\one_{\Hlow}.
 \label{eq:no-common}
\end{equation}
Thus, for $\bigcap_{i=1}^m R_i=\varnothing$, 
no nontrivial bulk operator can be commonly reconstructed from the regions \(R_i\).

For $X \equiv \bigcup_{i=1}^m R_i^c\neq \varnothing$,
the QFT detectability may mean
\begin{equation}
\Lcp(R_1^c,\ldots,R_m^c)' = \Rcp(X^c),
 \label{l2}
\end{equation}
where $X^c=\bigcap_{i=1}^m R_i$ and we have 
\begin{equation}
\Ccom(R_1,\ldots,R_m)  = \Rcp(\bigcap_{i=1}^m R_i).
 \label{c2}
\end{equation}
This is because, in the low-energy effective field theory, the only operators supported on $X$ that commute with all local operators on $X$ are trivial up to central, topological, or superselection data. These local operators include the smeared energy-momentum tensor on $X$, which generates local time evolution.
Thus, the condition \eqref{l2} is a consequence of a very basic fact: the existence of the energy-momentum tensor in QFT, although this fact is not respected in the quantum information theoretical models for holography in the literature.

\subsubsection{The three-region specialization}

Partition a time slice into three mutually disjoint regions $A_1,A_2,A_3$ and define\footnote{
Note that in AdS/CFT, the intersection of the three entanglement wedges of $R_i$ is not empty, at least for the AdS vacuum,
although $\bigcap_{i=1}^3 R_i=\varnothing$. Such an enlargement of the entanglement wedge and the bulk operators on it are the key to the paradoxes, which motivates the HQEC proposal in  \cite{Almheiri:2014lwa}.
The paradox itself is absent in the algebraic version of EWR, even without HQEC structure.}
\begin{equation}
 R_1=A_2\cup A_3,
 \qquad
 R_2=A_3\cup A_1,
 \qquad
 R_3=A_1\cup A_2.
 \label{eq:three-regions}
\end{equation}
Then $R_i^c=A_i$.  A $2$-out-of-$3$ logical operator, as in the HaPPY code \cite{Pastawski:2015qua}, would be a nontrivial element of
\begin{equation}
 \Rcp(A_2\cup A_3)
 \cap
 \Rcp(A_3\cup A_1)
 \cap
 \Rcp(A_1\cup A_2).
 \label{eq:two-out-of-three}
\end{equation}
However, as shown in \eqref{eq:no-common},
it is impossible to reconstruct the same bulk operator from three subregions, $R_i$.

Thus, the finite-dimensional statement isolates the kinematical core of the three-region argument in \cite{Sugishita:2024lee}.  
It does not use large $N$, gravity, an RT surface, or an entropy formula.  
The additional physical input is the usual local detectability of CFT/QFT observables, which was implemented concretely using the energy-momentum tensor and related low-energy observables in \cite{Sugishita:2024lee}.

\subsubsection{Two-region overlap version}
\label{subsec:two-region-overlap-version}

The same commutant argument gives a two-region form of the absence of
extra common reconstructions, which means the absence of the HQEC structure.  
Let $A$ and $B$ be two boundary regions and set
\begin{equation}
 C:=A\cap B .
 \label{eq:overlap-C}
\end{equation}
The HQEC claims the existence of a nontrivial
operator in
\begin{equation}
 \Rcp(A)\cap\Rcp(B)
 \label{eq:two-common}
\end{equation}
which is not reconstructible from the geometric overlap $C$.

Now, we apply \eqref{c2} for 
$R_1=A, R_2=B$; then we find 
\begin{equation}
\Rcp(A)\cap\Rcp(B)=\Rcp(A\cap B).
 \label{eq:no-extra-overlap-two-region}
\end{equation}
Therefore, if a common bulk field is reconstructible from both $A$ and $B$, its full
physical support must already be compatible with the geometric overlap.
Equation~\eqref{eq:no-extra-overlap-two-region} is the abstract form of the derivation given in Section~4.4 of \cite{Sugishita:2024lee}, in which a representative $O^{B}\in\mathcal A(B)$ is decomposed as $O^{B}=O^{B}_{A\cap B}+O^{B}_{\bar A}$, and commutation with the stress tensor supported in $\bar A$ forces the latter piece to act trivially on the low-energy sector.

Exact erasure codes such as HaPPY code \cite{Pastawski:2015qua} violate the local-detectability input in
\eqref{eq:LCI} and \eqref{l2}.  They are designed to contain a protected
logical algebra which is invisible to the erased-region local algebras, and this
protected commutant is precisely their error-correcting mechanism.
 The tensor network builds this protected commutant in by hand; it does not derive it from gravitational dynamics.

These findings imply that \eqref{eq:no-extra-overlap-two-region}—the non-existence of an HQEC-like structure derived from the basic properties of QFT—is actually a rather fundamental feature in holography.
Indeed, on the bulk side, gauge invariance via gravitational dressing plays a crucial role, and it appears to be linked to the physical consistency of holography.

\subsection{Algebraic EWR and gravitational dressing}
\label{sec:algebraic-EWR}

The discussion in the previous section concerns a specific extra property
which is often associated with holographic error correction: the same
nontrivial low-energy operator is assumed to have representatives in several
different boundary regions.  This is not what is meant by the algebraic version
of entanglement wedge reconstruction, as opposed to the claim in \cite{Harlow:2016vwg}.

In algebraic EWR, one assigns to each boundary region \(R\) the algebra
\(\Rcp(R)\) of physical operators which can be represented in \(R\).  In a
gravitational theory this statement includes the gravitational dressing.  Thus
an operator belongs to \(\Rcp(R)\) only if a complete gauge-invariant dressing
of that operator can be chosen with boundary support in \(R\).  Similarly,
\(\Rcp(R^c)\) consists of operators whose complete dressing can be chosen with
boundary support in \(R^c\).

With this interpretation, there is no reason to expect
\begin{equation}
        \Rcp(R)\vee\Rcp(R^c)
        =
        \End(\Hlow).
        \label{eq:not-expected-full-generation}
\end{equation}
Indeed, in gravity one generally expects the opposite.  A physical bulk
operator may require a gravitational dressing whose boundary support is not
contained entirely in \(R\) or entirely in \(R^c\).  Such an operator acts on
the low-energy Hilbert space, but it is not an element of either separately
defined regional algebra.  Therefore the algebra generated by the two regional
algebras can be a proper subalgebra,
\begin{equation}
        \Rcp(R)\vee\Rcp(R^c)
        \subsetneq
        \End(\Hlow).
        \label{eq:algebraic-nonfactorization}
\end{equation}
This is not an additional assumption.  It is the usual non-factorization of
gauge-invariant operator algebras in the presence of Gauss-law constraints,
here applied to gravitational dressing.

The statement \eqref{eq:algebraic-nonfactorization} should be distinguished
from the property discussed in the previous section.  Equation
\eqref{eq:algebraic-nonfactorization} says that the two separately defined
regional algebras, those associated with \(R\) and \(R^c\), need not generate
all operators on \(\Hlow\).  By contrast, the HQEC-type property is an
intersection statement.  It would require a nontrivial operator \(W\) such that
\begin{equation}
        W
        \in
        \Rcp(R_1)\cap\Rcp(R_2)\cap\cdots ,
        \qquad
        W\notin \mathbb C\one_{\Hlow},
        \label{eq:HQEC-intersection}
\end{equation}
for several distinct boundary regions \(R_i\).  In words, the same operator on
\(\Hlow\) would have several independent regional representations.

Thus, there are two different statements:
\begin{align}
        \Rcp(R)\vee\Rcp(R^c)
        &\subsetneq
        \End(\Hlow),
        \label{eq:generation-statement}
        \\
        \Rcp(R_1)\cap\Rcp(R_2)\cap\cdots
        &\supsetneq
        \mathbb C\one_{\Hlow}.
        \label{eq:intersection-statement}
\end{align}
The first statement is compatible with, and in gravity expected from, the
algebraic interpretation of EWR.  The second statement is the additional
HQEC-type property.  The claim of this paper is that the second statement is
not realized by ordinary finite \(N\) low-energy CFT/SUGRA operators.

This distinction is important.  Algebraic EWR assigns to each boundary region
the physical algebra of operators whose complete gravitational dressing can be
placed in that region.  It does not say that the algebras associated with
\(R\) and \(R^c\) are two tensor factors whose product gives all of
\(\End(\Hlow)\).  The failure of such a tensor-factor description is a standard
feature of gauge theories and gravity, not a problem for algebraic EWR.  What
fails is the stronger HQEC picture in which a protected nontrivial operator is
simultaneously reconstructible from several different boundary regions. 

\paragraph{Remark on complementary recovery.}
The algebraic form of complementary recovery is the commutant relation
\begin{equation}
        \Rcp(R)'=\Rcp(R^c),
        \label{eq:algebraic-complementary-recovery}
\end{equation}
where the commutant is taken in \(\End(\Hlow)\), with the usual qualification
that possible center sectors are treated block by block.
Note that the property \eqref{eq:algebraic-complementary-recovery} itself can exist without QEC structure and
this statement
should not be confused with the HQEC property discussed above.  Equation
\eqref{eq:algebraic-complementary-recovery} relates two complementary regional
algebras.  It does not say that a nontrivial operator belongs simultaneously
to several different regional algebras.\footnote{
Harlow formulates the algebraic RT formula in the language of operator
algebra quantum error correction \cite{Harlow:2016vwg}.  However, the part of the argument
that is relevant for the entropy formula only needs an algebraic
relation between the region \(R\) and its complement \(R^c\) \eqref{eq:algebraic-complementary-recovery}, such as a
pair of complementary algebras \(M_R\) and \(M_R'\).  By itself, this is
not the extra common reconstruction structure that one normally
associates with holographic error correction.  The latter would require
the same logical operator to be reconstructible from several different
boundary regions.
}
We stress that the commutant relation
\eqref{eq:algebraic-complementary-recovery}
should not be interpreted as a quantum-error-correcting structure.  It is a
statement about complementary subalgebras of the low-energy theory.  A
correctable-algebra structure would require an independently defined logical
algebra whose nontrivial elements admit code-preserving representatives after
erasure of complementary boundary degrees of freedom.  This is precisely the
redundant logical-operator structure of the ADH/HaPPY interpretation.  The
point of the present argument is that such a structure is absent for ordinary
bulk-field operators.

In a gravitational theory,
\eqref{eq:algebraic-complementary-recovery} may mean that the union of the entanglement wedges of $R$ and $R^c$ is the whole space. 
Here, we will not discuss whether or not this is realized in AdS/CFT in general.

\section{Why the strict \texorpdfstring{$N=\infty$}{N=infinity} limit is not evidence for HQEC}
\label{sec:gff}

In the strict $N=\infty$ limit, the holographic CFT sector relevant to a fixed semiclassical background is a generalized free field (GFF) theory.  It is equivalent to a free bulk theory, with the radial direction encoded as an additional mode label.  Within this theory the space conventionally called the code subspace is the physical Hilbert space of the GFF sector itself \cite{Sugishita:2023wjm, Sugishita:2024lee}.  Thus
\begin{equation}
 \Hlow^{N=\infty}=\Hphys^{\rm GFF}
 \label{eq:gff-code-physical}
\end{equation}
for the theory under discussion.

An equality between global and Rindler free-field bulk constructions at this order \cite{Hamilton:2006az} is therefore an identity or equivalence inside one free theory.  It is not the protection of a smaller interacting logical sector against erasure of part of a larger physical system.  
Moreover, the GFF sector is not an ordinary local QFT (on the boundary).  It has no
stress tensor within the generalized-free sector, and it does not satisfy
the time-slice axiom.  Consequently, local generalized-free operators do
not generate the same algebraic structure as the local operators of the well-defined QFT, i.e. 
finite \(N\) CFT.  In particular, the finite \(N\) stress tensor provides
local probes which can distinguish reconstructions that are indistinguishable
in the strict GFF approximation.
The apparent local invisibility that makes a QEC interpretation possible is already a special property of the $N=\infty$ limit, not a property of the finite $N$ CFT \cite{Sugishita:2023wjm, Sugishita:2024lee}.

Let $g$ denote the first interaction order; depending on conventions, $g$ may scale as $1/N$, a power of $G_N^{1/2}$, or the first nonzero normalized three-point-function coefficient.  Write two reconstructions as
\begin{align}
 O_A(g)&=O^{(0)}+gO_A^{(1)}+O(g^2),\\
 O_B(g)&=O^{(0)}+gO_B^{(1)}+O(g^2).
 \label{eq:pert-reconstructions}
\end{align}
HQEC, through the first gravitational order, would require
\begin{equation}
 P \bigl(O_A(g)-O_B(g)\bigr)P=O(g^2).
 \label{eq:needed-order}
\end{equation}
The results of \cite{Sugishita:2023wjm, Sugishita:2024lee} show that this fails, in the following precise form.  Let $O^{\delta}:=O_A(g)-O_B(g)$.  First, the difference annihilates the vacuum,
\begin{equation}
 O^{\delta}\,\ket{0}=0, \,\, \bra{0} O^{\delta} =0,
 \label{eq:vac-annihilation}
\end{equation}
i.e.\ all two-point functions of $O^{\delta}$ with primary operators vanish.  
The difference between the two reconstructions is therefore invisible at the level of two-point functions; this is why checks performed in the strict GFF/two-point approximation suggest an exact code.  
The mismatch first appears in three-point functions containing $O^{\delta}$,
\begin{equation}
 \bra{0}\, O_1 
 \; O^{\delta}\; O_2 \,\ket{0}
 =c\,g+O(g^2),
 \label{eq:first-order-difference}
\end{equation}
where $O_1, O_2$ are low-energy probes and $ c\neq0$ for some $O_1, O_2$.  Equivalently, with the unnormalized CFT stress tensor, the mismatch is of order $g^{0}$.

Moreover, the mismatch \eqref{eq:first-order-difference} cannot be removed by any admissible redefinition of the regional representatives, i.e., the $1/N$ corrections to the bulk reconstruction.  
The separating functional used in \cite{Sugishita:2024lee} is the double commutator with the energy density smeared over a region $\Omega$ contained in the complement $\bar A$,
\begin{equation}
 \bigl\langle 0\big|\,
 \bigl[\,\widetilde O_A,\bigl[\,T_{00}(\Omega),\widetilde O_A\,\bigr]\bigr]
 \,\big|0\bigr\rangle,
 \label{eq:double-commutator}
\end{equation}
which vanishes identically for every operator $\widetilde O_A$ supported in $A$, by microcausality alone.
For the global reconstruction $\widetilde O_G$ which is not supported in $A$, the same double commutator 
is instead fixed by rotational invariance to be
 $(2/V)\,\langle 0|\widetilde O_G\,H\,\widetilde O_G|0\rangle>0$ at order $g^{0}$, and this evaluation uses low-energy states only.  Since a $1/N$ correction to a representative cannot contribute at order $g^{0}$, no choice of representative in $\mathcal A(A)$, at any order in $g$, restores the equality \eqref{eq:needed-order}.  

We stress that the probe which separates the two reconstructions is precisely the energy-momentum tensor: the same observable whose local detectability is the physical input of Section~\ref{subsec:local-qft-detectability}, and precisely the observable that HaPPY-type models lack.  In the bulk language, \eqref{eq:double-commutator} states that the $A$-dressed and the $B$-dressed operators carry different gravitational field configurations, and the (boundary) energy-momentum tensor detects this difference.  The CFT computation of \cite{Sugishita:2023wjm, Sugishita:2024lee} and the finite-dimensional lemma of Section~\ref{sec:finite-dimensional} are in this sense two forms of a single statement.

The equality \eqref{eq:needed-order} is therefore lost precisely at the first order at which gravity or interaction becomes visible.

Calling \eqref{eq:first-order-difference} an ``approximate code with error $O(g)$'' is mathematically allowed, but it does not establish an HQEC structure through order $g$.  It says only that the free-theory relation is recovered when the first gravitational effect is neglected.

\section{AdS/CFT interpretation: EWR without HQEC}
\label{s5}

The finite-dimensional discussion above should not be read as a new
derivation of the results in \cite{Sugishita:2023wjm, Sugishita:2024lee}.  Rather, it isolates the
quantum-information structure which those results expose in holographic
CFT.  The abstract statement is
\[
  \text{common reconstruction}
  \quad\Longrightarrow\quad
  \text{local invisibility}.
\]
If the same low-energy operator has code-preserving representatives in
several boundary regions, then it must commute with the corresponding
local observables in the complementary regions.  HaPPY-type codes make
this possible by construction: the erased-region algebras have a
nontrivial protected commutant.

The finite \(N\) holographic CFT behaves differently.  The local CFT
observables, in particular the energy-momentum tensor and other low-energy
single-trace observables, detect ordinary low-energy bulk excitations.
This fact was used to show that global and subregion
reconstructions which agree in the strict free limit become
distinguishable at the first nontrivial \(1/N\) order.  In the bulk
language, this is the statement that gravitationally dressed operators
with different boundary support are different physical operators.  Thus
the common-logical-operator structure required by HQEC is absent for
ordinary interacting bulk fields.

What remains is EWR without HQEC.  For each boundary region \(A\), the
code-preserving low-energy algebra $\Rcp(A)$ of suitably smeared CFT operators can be given a
semiclassical bulk interpretation, when such an interpretation is
available.  It is interpreted as the algebra of gauge-invariant bulk
operators whose full gravitational dressing is compatible with the
region \(A\).  A different region \(B\) gives a different algebra
\(\Rcp(B)\), with a different dressing prescription.  The two
descriptions may agree in the free limit, but they are not required to be
representations of a logical operator.

Thus the result of removing HQEC is not the absence of subregion
reconstruction.  It is the replacement of the error-correcting statement
``one logical operator, many boundary representatives'' by the
region-dependent statement ``one boundary region, one dressed algebraic
interpretation.'' 
Thus, the finite \(N\) remnant of subregion reconstruction is not HQEC but subregion complementarity \cite{Sugishita:2023wjm}: each boundary region admits its own
region-adapted bulk description, and different descriptions need not be
representations of the same physical operator.

In the discussion up to this point, no assumption has been made regarding the order up to which the EWR remains valid in the $1/N$ expansion.
Suppose, optimistically, that a refined JLMS statement including the correct QES/area terms holds to all orders in \(1/N\).\footnote{
JLMS was derived only to leading order \cite{Jafferis:2015del, Faulkner:2013ana}; the QES prescription was conjectured to hold at arbitrary orders \cite{Engelhardt:2014gca}; the optimistic assumption is that the refined relation holds to all orders.
Furthermore, the entanglement entropy in the replica trick may yield an algebraic version of the entanglement entropy, as claimed in \cite{Ghosh:2015iwa} for the algebraic version defined in \cite{Casini:2013rba} in a gauge theory.}
This is naturally consistent with EWR without HQEC: each boundary region has
its own region-dependent algebra, with its own gravitationally dressed
bulk interpretation.  
It still would not imply the HQEC identity that
the same finite \(N\) physical operator has representatives in several regions.

Finally, we distinguish the issue discussed in this paper from a separate
finite \(N\) breakdown of the large \(N\) reconstruction expansion. 
Throughout we have worked within the large $N$ expansion; at finite $N$, however, there are quantities for which the expansion itself breaks down.
In fact, reference \cite{Terashima:2025shl} shows that the subregion reconstruction of a bulk local operator fails when the energy scale of the smearing function reaches the order of $\ln N$.\footnote{
This breakdown can be avoided by removing the components corresponding to the horizon-to-horizon null geodesics \cite{Bousso:2012mh}, which is related to the discussion of the no-common reconstruction \cite{Sugishita:2022ldv}.} 
This implies that bulk reconstruction in the presence of a horizon—such as when considering a subregion—causes issues within the large $N$ expansion. A similar breakdown of the large $N$ expansion has also been discussed in black hole backgrounds, which is related to the information paradox \cite{Terashima:2026fix, Chikazawa:2026iro, Terashima:2025tct, Iizuka:2013kma}.
While it is important to discuss these finite $N$ effects specific to entanglement wedge reconstruction (EWR), we leave this to future work.

\section*{Acknowledgements}

The author would like to thank H. Kanda and S. Sugishita for their useful discussions.
This work was supported by JSPS KAKENHI Grant Number 	24K07048.

\section*{Declaration of AI-assisted tools}

During the preparation of this manuscript, AI-assisted tools were used
for language editing and for preliminary consistency checks of some
intermediate formulae.  All scientific content, calculations, and conclusions
were independently verified by the author.

\hspace{1cm}

\appendix

\section{Why exact reconstruction implies code preservation}
\label{app:code-preservation}

Let $V:\Hlogical\to\Hfull$ be an isometric encoding and let $P=VV^\dagger$.  Suppose a physical operator $O$ represents a logical operator $o$ in the usual exact sense,
\begin{equation}
 OV=Vo,
 \qquad
 O^\dagger V=Vo^\dagger.
 \label{eq:intertwining}
\end{equation}
Taking the adjoint of the second relation gives
\begin{equation}
 V^\dagger O=oV^\dagger.
\end{equation}
Therefore
\begin{align}
 OP&=OVV^\dagger=VoV^\dagger,\label{eq:OP}\\
 PO&=VV^\dagger O=VoV^\dagger.\label{eq:PO}
\end{align}
Hence
\begin{equation}
 [O,P]=0.
 \label{eq:exact-implies-preserve}
\end{equation}
Thus the code-preserving requirement is not an extra restriction on exact $*$-algebra reconstruction.  It is what distinguishes exact physical representatives from arbitrary compressions.

\paragraph{Projection and locality}
\label{app:compression}
For generic $a,b\in\End(\Hfull)$, even if $[a,b]=0$, their projections need not commute:
\begin{equation}
 [PaP,PbP]
 =Pb(\one-P)aP-Pa(\one-P)bP.
 \label{eq:compression-nonlocal}
\end{equation}
If $[a,P]=0$, however,
\begin{equation}
 [PaP,PbP]
 =P[a,b]P.
 \label{eq:compression-local-preserve}
\end{equation}
This identity is the reason code preservation is essential in Lemma~\ref{lem:commutant-bound}.  The reconstructed representative must preserve the low-energy sector.  The complementary observable must also preserve the low-energy sector in the form used in the lemma.

\hspace{1cm}

\bibliographystyle{utphys}
\bibliography{main202506}

\end{document}